\documentclass[9pt, conference]{IEEEtran}
\IEEEoverridecommandlockouts
\usepackage{cite}
\usepackage{amsmath,amssymb,amsfonts}
\usepackage{graphicx}
\usepackage{textcomp}
\def\BibTeX{{\rm B\kern-.05em{\sc i\kern-.025em b}\kern-.08em
    T\kern-.1667em\lower.7ex\hbox{E}\kern-.125emX}}

\usepackage{algorithm}
\usepackage{algpseudocode}
\usepackage{array}
\usepackage[caption=false,font=footnotesize,labelfont=rm,textfont=rm]{subfig}
\usepackage{textcomp}
\usepackage{hyperref}
\usepackage{stfloats}
\usepackage{url}
\usepackage{verbatim}
\usepackage{tablefootnote}
\usepackage{color, colortbl}
\usepackage[capitalize,noabbrev]{cleveref}
\usepackage{caption2}
\usepackage{booktabs}
\usepackage{threeparttable}
\usepackage[x11names]{xcolor}
\usepackage{subfloat}
\usepackage{multirow}

\usepackage{xspace}
\def\eg{\emph{e.g.}\xspace} 

\def\ie{\emph{i.e.}\xspace}

\definecolor{darkred}{rgb}{0.7,0,0}
\definecolor{darkgreen}{rgb}{0,0.46,0}
\definecolor{purple}{rgb}{0.6,0,0.5}
\definecolor{cholocate}{HTML}{d2691e}
\definecolor{slateblue}{HTML}{6a5acd}

\newcommand{\fin}{\color{black}}

\newcommand{\nameFramework}{GN-FT\xspace}

\begin{document}

\title{Gradient Norm-based Fine-Tuning for Backdoor Defense in Automatic Speech Recognition
}



\author{
  Nanjun Zhou\textsuperscript{1,2,*}, Weilin Lin\textsuperscript{1,*}\thanks{* Equal Contribution.}, Li Liu\textsuperscript{1,$\dagger$} \thanks{$\dagger$ Corresponds to Li Liu (\href{mailto:avrillliu@hkust-gz.edu.cn}{avrillliu@hkust-gz.edu.cn}).}\\
  \textsuperscript{1}\textit{The Hong Kong University of Science and Technology (Guangzhou), Guangzhou, China}\\
  \textsuperscript{2}\textit{South China University of Technology, Guangzhou, China}}

\maketitle

\begin{abstract}
Backdoor attacks have posed a significant threat to the security of deep neural networks (DNNs). Despite considerable strides in developing defenses against backdoor attacks in the visual domain, the specialized defenses for the audio domain remain empty. Furthermore, the defenses adapted from the visual to audio domain demonstrate limited effectiveness. To fill this gap, we propose \textit{\textbf{G}radient Norm-based \textbf{F}ine-\textbf{T}uning (\nameFramework)}, a novel defense strategy against the attacks in the audio domain, based on the observation from the corresponding backdoored models. Specifically, we first empirically find that the backdoored neurons exhibit greater gradient values compared to other neurons, while clean neurons stay the lowest. 
On this basis, we fine-tune the backdoored model by incorporating the gradient norm regularization, aiming to weaken and reduce the backdoored neurons. We further approximate the loss computation for lower implementation costs. 
Extensive experiments on two speech recognition datasets across five models demonstrate the superior performance of our proposed method. To the best of our knowledge, this work is the first specialized and effective defense against backdoor attacks in the audio domain. 
\end{abstract}


\section{Introduction}
\label{Introduction}

In recent years, deep neural networks (DNNs) have seen extensive application across a wide range of fields, including face recognition \cite{taigman2014deepface, parmar2014face, ibrahim2011study}, autonomous driving \cite{caesar2020nuscenes,yurtsever2020survey} in the visual domain, and automatic speech recognition \cite{warden2018speech, malik2021automatic} in the audio domain. However, with advancements in technology, backdoor attacks \cite{gu2019badnets} have emerged as a severe security concern, threatening the safety of DNNs. In backdoor attacks, attackers inject a specific \textit{trigger} pattern into a portion of training dataset to poison the data. Models trained on such poisoned datasets, known as backdoored models, behave normally when presented with clean data. Conversely, they maliciously misclassify data that contains the trigger pattern to a predefined target label, which is termed as backdoored effect. To avoid this effect, solutions on either the poisoned-input detection (data-level) or the backdoored-model repairing (model-level) are necessary~\cite{yan2023backdoor}. \fin

Up to now, numerous studies have been dedicated to developing defenses against backdoor attacks, which have achieved significant results~\cite{liu2018fine, wu2021adversarial, li2021anti, huang2022backdoor, li2023reconstructive, wu2023defenses}. However, these defense methods are mostly designed for the visual domain, and no specialized defense is proposed against the backdoor attacks in the audio domain. Due to the different characteristics between the two domains, \eg, audio signals own larger information density as spectrum format compared to the RGB images, the existing defense methods adapted from the visual domain~\cite{liu2018fine, gao2019strip, ma2022beatrix} demonstrate limited performance against the audio backdoor attacks.
As illustrated in Table~\ref{table_result-SCD10} and Table~\ref{table_result-SCD30}, the model-level defense adapted from the visual domain, Fine-Pruning (FP)~\cite{liu2018fine}, fails completely in the audio domain. Therefore, in this work, we aim to design the first defense method specifically targeted at audio backdoor attacks.

\begin{figure}[ht]
\setlength{\abovecaptionskip}{3pt}
\begin{minipage}[b]{1.0\linewidth}
\centering
\centerline{\includegraphics[width=8cm]{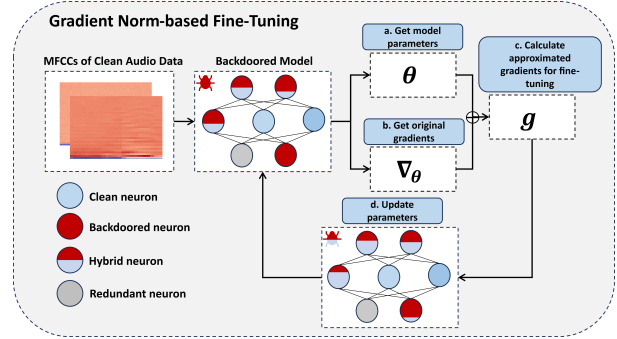}}
\end{minipage}
\caption{Overview of our proposed method (\nameFramework). }
\label{fig_pipeline}
\end{figure}

\begin{figure}[t]
\setlength{\abovecaptionskip}{3pt}
\begin{minipage}[b]{1.0\linewidth}
 \centering
\centerline{\includegraphics[width=8.8cm]{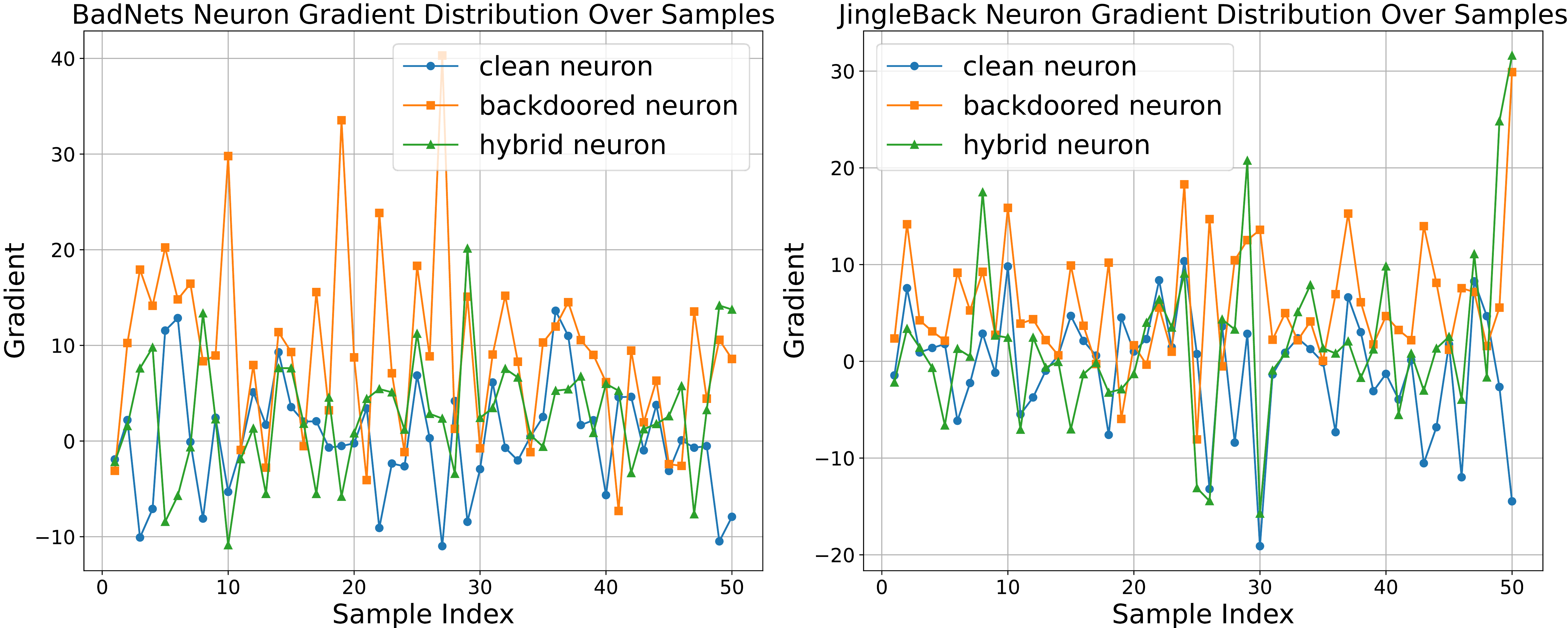}}
\end{minipage}
\caption{Illustration of gradients of different neurons over 50 clean samples. We used Audio BadNets~\cite{gu2019badnets} and JingleBack \cite{koffas2023going} on ResNet \cite{he2016deep} for illustrations. For most clean inputs, backdoored neurons and hybrid neurons exhibit larger gradients, while clean neurons show smaller gradients. }
\label{fig_grad}
\vspace{-1.5em}
\end{figure}

To investigate the characteristics of the backdoored models in the audio domain, \ie, \textit{audio-backdoored models}, we split its neurons into different types as in \cite{li2024magnitude} and further observe their learning behaviors. Specifically, the neurons are categorized into clean neurons, backdoored neurons, hybrid neurons, and redundant neurons according to their loss changes on both clean and backdoor tasks\footnote{Clean task represents the normal classification task on clean samples. Similarly, backdoor task indicates the task on poisoned samples. }. Note that backdoored neurons and hybrid neurons are the primary contributors to the backdoor task, and our goal is to weaken their functionality. The gradients with clean inputs on different neuron types are shown in Fig.~\ref{fig_grad}, we can observe that for most clean inputs, backdoored neurons and hybrid neurons exhibit larger gradient values than clean neurons, which solely contribute to clean task. 
     
Based on this observation, we propose \textbf{G}radient \textbf{N}orm-based \textbf{F}ine-\textbf{T}uning (\textbf{\nameFramework}), where a gradient norm regularization term is added to the original loss function. By doing so, 
the learning process attempts to suppress the high-gradient backdoored neurons and hybrid neurons, resulting in a repaired clean model after fine-tuning. 
Considering computational efficiency, we adopt the approximation scheme introduced in \cite{zhao2022penalizing}. Extensive experiments demonstrate that our method significantly outperforms FP in terms of defense effectiveness.

In summary, the main contributions of this work are threefold: \textbf{1) }We observe that the backdoored neurons in the audio-backdoored models exhibit greater gradient values than others. \textbf{2) }We propose a gradient-regularized fine-tuning method to effectively mitigate backdoored effect, marking the first specialized defense technique for backdoor attacks in the audio domain. \textbf{3) }Extensive experiments across two datasets, five models, and seven attacks, show that our proposed method consistently achieves state-of-the-art performance. 

\section{Related work}

\noindent \textbf{Backdoor Attacks. }In backdoor attacks \cite{gu2019badnets}, an attacker injects a specific pattern, known as \textit{trigger}, into a portion of the training data and assigns these samples a target label. The resulting backdoored model performs normally on clean data but misclassifies inputs with the trigger to the target label. Most backdoor attack techniques \cite{gu2019badnets, chen2017targeted, nguyen2020input, nguyen2021wanet, wang2022bppattack, wu2023attacks} are designed for the visual domain, among which classic examples include BadNets \cite{gu2019badnets} and Blended \cite{chen2017targeted}. In the audio domain, Ultrasonic attack \cite{koffas2022can} is a representative method for automatic speech recognition tasks, where the attacker uses an ultrasonic signal as the backdoor trigger. To enable attacks in a physical scenario, naturally occurring sounds are chosen as triggers in DABA \cite{liu2022opportunistic}. Various audio-specific methods \cite{koffas2023going, cai2024towards} have also been devised to increase the stealthiness of attacks. Recently, a stealthy attack FlowMur \cite{lan2024flowmur} was introduced, where it trains a model to generate triggers while ensuring consistency between the target label and ground truth.

\noindent \textbf{Backdoor Defenses. }According to \cite{yan2023backdoor}, backdoor defenses can be categorized into data-level and model-level approaches. Data-level defenses aim to identify and remove poisoned data from the dataset, while model-level defenses attempt to mitigate backdoored effect in a well-trained backdoored model using a small amount of clean data. In the audio domain, existing backdoor defenses are all adaptations from the visual domain and are primarily data-level \cite{gao2019strip, ma2022beatrix}. FP \cite{liu2018fine} is the only adapted model-level defense for audio-backdoored models, which prunes neurons with low activation on clean data and then fine-tunes the pruned model. However, FP fails to effectively defend against most audio backdoor attacks. In this work, we address this issue by proposing a gradient-regularized fine-tuning technique from the model-level perspective, which is the first specialized defense for the audio-backdoored models. 



\section{Proposed Method}
\subsection{Problem Formulation}  
\noindent \textbf{Threat Models. }We assume that the attackers have full access to the training set, and they poison it by injecting a trigger into a small amount of randomly selected samples, indicated by the \textit{poisoning ratio}. The attackers aim to train the model with the poisoned training set so that it misclassifies poisoned data as the target label $y_t$, while functioning normally on clean data. We denote the model as $F$ with $L$ layers, where $f^{(i)}$ parameterized as $\boldsymbol{\theta}^{(i)}$ denotes $i$-th layer of the model. Considering the convolutional layer, the weights of neurons in the $i$-th layer can be denoted as $\{\boldsymbol{\theta}^{(i, j)} \in \mathbb{R}^{c_{i-1} \times h \times w}\}_{1\leq j\leq c_i}$, where $c_i$, $h$ and $w$ represent the number of neurons in $f^{(i)}$, the height and width of the convolutional kernel, respectively.

\noindent \textbf{Defense Setting. }The defender aims to eliminate the backdoored effect while preserving the model performance on clean data. Following the previous model-level defense settings \cite{liu2018fine}, we assume that only 5\% of clean data is accessible to the defender for conducting defense, denoted as $\mathcal{D}_c$. 

\subsection{Observations on Audio-Backdoored Models}
\label{Observations}
\noindent \textbf{Classification of Neurons in Backdoored Models. }In line with the neuron types defined in \cite{li2024magnitude}, we categorize the neurons in backdoored models based on pruning and loss change, where pruning \textit{j}-th neuron in the \textit{i}-th layer of the model means setting $\boldsymbol{\theta}^{(i, j)}$ to 0. The loss change of a neuron is defined as the difference between the loss values after and before pruning for the same inputs. To be specific, \textit{Clean Loss Change} (CLC) and \textit{Backdoor Loss Change} (BLC) can be formulated as: 
\begin{align}
  \text{CLC}(\boldsymbol{\theta}, i, j) &= \mathbb{E}_{(\boldsymbol{x}, y) \in \mathcal{D}_c} \bigl[ \mathcal{L}(F(\boldsymbol{x}; \boldsymbol{\theta} \mid \boldsymbol{\theta}^{(i, j)}=0), y) \notag \\
  &\quad - \mathcal{L}(F(\boldsymbol{x}; \boldsymbol{\theta}), y) \bigr]  \\
  \text{BLC}(\boldsymbol{\theta}, i, j) &= \mathbb{E}_{(\boldsymbol{x}, *) \in \mathcal{D}_c, y_t} \bigl[ \mathcal{L}(F(\delta(\boldsymbol{x}); \boldsymbol{\theta} \mid \boldsymbol{\theta}^{(i, j)}=0), y_t) \notag \\
  &\quad - \mathcal{L}(F(\delta(\boldsymbol{x}); \boldsymbol{\theta}), y_t) \bigr] 
\end{align} 
where $\mathcal{D}_c$ is the given clean data for defense; $\delta(\cdot)$  is the poisoning function of the attack method; $y_t$ is a predefined target label; and $\mathcal{L}(\cdot)$ is the \textit{cross-entropy loss}. Note that a larger value of CLC (or BLC) represents a larger contribution of the current neuron to the clean task (or backdoor task). Therefore, we can adopt it to classify the neurons into different types.

\begin{figure}[t]
\setlength{\abovecaptionskip}{3pt}
\begin{minipage}[b]{1.0\linewidth}
 \centering
 \centerline{\includegraphics[width=5cm]{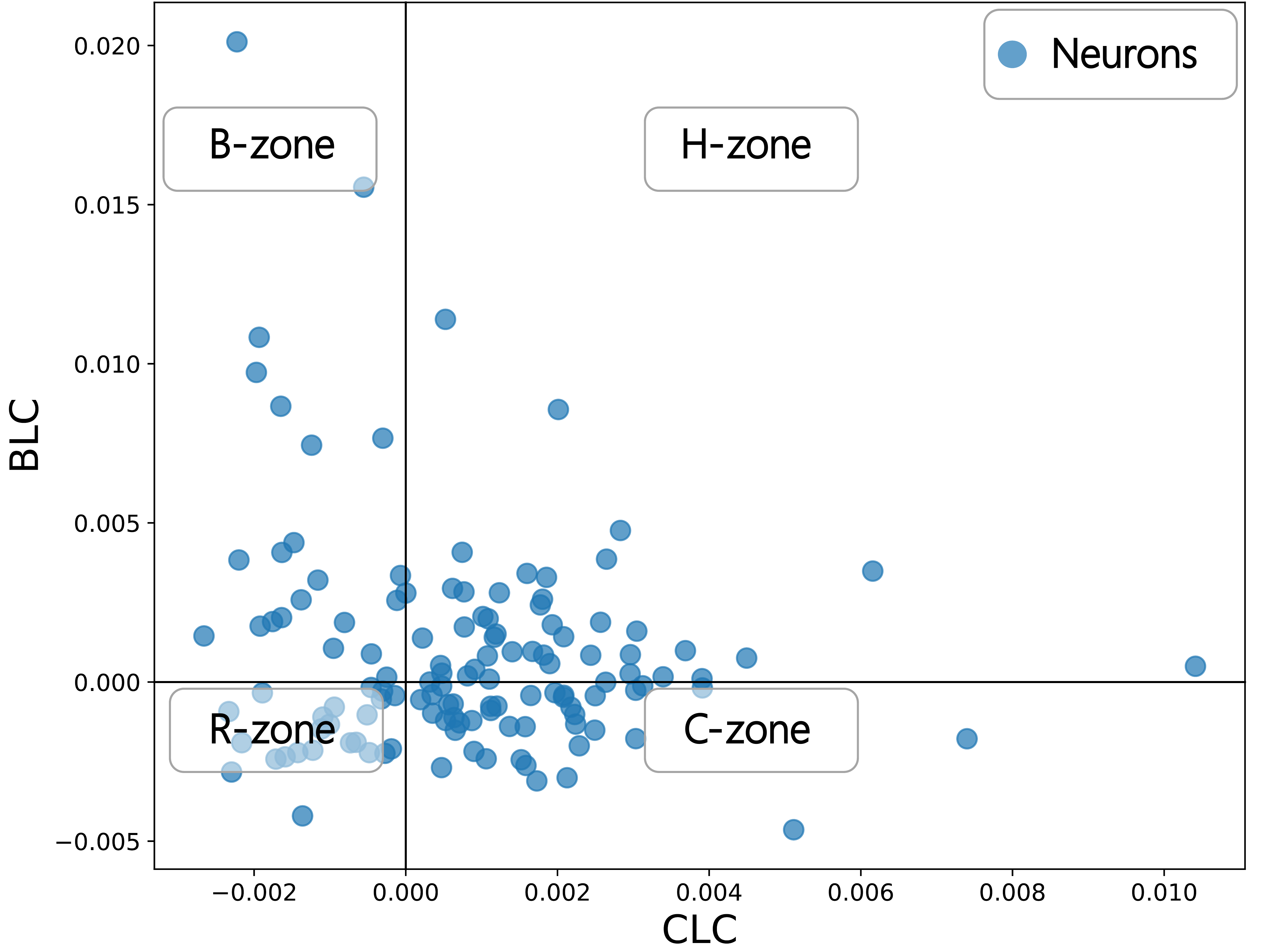}}
\end{minipage}
\caption{A scatter plot showing the BLC and CLC values for neurons in the last two convolutional layers of an audio-backdoored model attacked by Audio BadNets \cite{gu2019badnets}.
\textbf{C-zone}: Clean Zone; \textbf{B-zone}: Backdoor Zone; \textbf{H-zone}: Hybrid Zone; \textbf{R-zone}: Redundant Zone.
}
\label{fig_example}
\vspace{-1.5em}
\end{figure}

Using this definition, we can obtain the CLCs and BLCs of all neurons within the audio-backdoored model, as illustrated in Fig. \ref{fig_example}. We divide the plot into four zones based on the zero values of CLC and BLC: The \textit{Clean Zone} (C-zone) contains neurons with positive CLCs and negative BLCs, indicating their contribution to the clean task while potentially suppressing the backdoor task. The \textit{Backdoor Zone} (B-zone) is characterized by neurons with positive BLCs and negative CLCs, suggesting a specific contribution to the backdoor task, potentially at the expense of the clean task. In the \textit{Hybrid Zone} (H-zone), neurons have both positive CLCs and BLCs, meaning they could contribute to both tasks. Finally, the \textit{Redundant Zone} (R-zone) contains neurons that do not contribute to either task. Based on these four zones, we can further classify neurons into \textbf{clean neurons}, \textbf{backdoored neurons}, \textbf{hybrid neurons}, and \textbf{redundant neurons}, respectively. Intuitively, our goal is to suppress backdoored neurons and hybrid neurons for backdoor mitigation.

\noindent \textbf{Suggestion Given by the Observation. }As illustrated in Section \ref{Introduction}, in the audio-backdoored models, \textbf{backdoored neurons and hybrid neurons tend to exhibit larger gradients on most clean inputs, while clean neurons stay the smallest.}
Therefore, it suggests penalizing the high gradient norm during fine-tuning to repair these two kinds of neurons, where the clean neurons are less modified due to their small-gradient characteristic. 




\subsection{Gradient Norm-based Fine-Tuning}
Based on the observations, we propose \textit{Gradient Norm-based Fine-Tuning} to penalize the high gradients from backdoor neurons and hybrid neurons. An overview of the method is shown in Fig. \ref{fig_pipeline}. An $L_2$ norm of the gradients is added as a regularization term to the fine-tuning loss function, as shown below:
\begin{equation}
\label{equ:gradient_loss}
\mathcal{L}(\boldsymbol{\theta})=\mathcal{L}_{c e}(\boldsymbol{\theta})+\lambda \cdot\left\|\nabla_{\boldsymbol{\theta}} \mathcal{L}_{c e}(\boldsymbol{\theta})\right\|_2,
\end{equation}
where $\mathcal{L}_{c e}(\cdot)$ is the original cross-entropy loss,  $\left\|\nabla_{\boldsymbol{\theta}} \mathcal{L}_{c e}(\boldsymbol{\theta})\right\|_2$ corresponds to the $L_2$ norm of the gradients of the model, and $\lambda$ is a trade-off coefficient to control the strength of penalization. 
During the fine-tuning process, we aim to minimize this loss function using the available clean set $D_c$.  The object function is formulated as:
\begin{equation}
\min _{\boldsymbol{\theta}} \mathbb{E}_{(\boldsymbol{x}, y) \in \mathcal{D}_c}[\mathcal{L}(F(\boldsymbol{x} ; \boldsymbol{\theta}), y)].
\end{equation}

However, direct optimization on equation~\eqref{equ:gradient_loss} involves calculating a Hessian matrix with $O(n^2)$ time and space complexities, which is infeasible. 
Inspired by the approximation scheme in \cite{zhao2022penalizing}, we choose to simplify it similarly using Taylor expansion and additional optimization steps. 
It can be formulated as:  
\begin{equation}
\label{equ:appro}
\scalebox{0.9}{$
\begin{aligned}
\nabla_{\boldsymbol{\theta}} \mathcal{L}(\boldsymbol{\theta})
& =\nabla_{\boldsymbol{\theta}} \mathcal{L}_{c e}(\boldsymbol{\theta}) + \nabla_{\boldsymbol{\theta}}( \lambda \cdot\left\|\nabla_{\boldsymbol{\theta}} \mathcal{L}_{c e}(\boldsymbol{\theta})\right\|_2 )\\
& =\nabla_{\boldsymbol{\theta}} \mathcal{L}_{c e}(\boldsymbol{\theta})+\lambda \cdot \nabla_{\boldsymbol{\theta}}^2 \mathcal{L}_{c e}(\boldsymbol{\theta}) \frac{\nabla_{\boldsymbol{\theta}} \mathcal{L}_{c e}(\boldsymbol{\theta})}{\left\|\nabla_{\boldsymbol{\theta}} \mathcal{L}_{c e}(\boldsymbol{\theta})\right\|_2} \\
& \approx \nabla_{\boldsymbol{\theta}} \mathcal{L}_{c e}(\boldsymbol{\theta})+\frac{\lambda}{r} \cdot(\nabla_{\boldsymbol{\theta}} \mathcal{L}_{c e}(\boldsymbol{\theta}+r \frac{\nabla_{\boldsymbol{\theta}} \mathcal{L}_{c e}(\boldsymbol{\theta})}{\left\|\nabla_{\boldsymbol{\theta}} \mathcal{L}_{c e}(\boldsymbol{\theta})\right\|_2})-\nabla_{\boldsymbol{\theta}} \mathcal{L}_{c e}(\boldsymbol{\theta})) \\
& =(1-\alpha) \nabla_{\boldsymbol{\theta}} \mathcal{L}_{c e}(\boldsymbol{\theta})+\alpha \nabla_{\boldsymbol{\theta}} \mathcal{L}_{c e}(\boldsymbol{\theta}+r \frac{\nabla_{\boldsymbol{\theta}} \mathcal{L}_{c e}(\boldsymbol{\theta})}{\left\|\nabla_{\boldsymbol{\theta}} \mathcal{L}_{c e}(\boldsymbol{\theta})\right\|_2}),
\end{aligned}$}
\end{equation} 
where $r$ is for appropriating the Hessian multiplication, and $\alpha=\frac{\lambda}{r}$ is used for trade-off.
In the practical defense process, we conduct an additional optimization step to approximate the second term in equation~\eqref{equ:appro}, aiming to avoid the Hessian computation:
\begin{equation}
    \label{equ:loss_appro_second}  
    \nabla_{\boldsymbol{\theta}} \mathcal{L}_{c e}(\boldsymbol{\theta}+r \frac{\nabla_{\boldsymbol{\theta}} \mathcal{L}_{c e}(\boldsymbol{\theta})}{\left\|\nabla_{\boldsymbol{\theta}} \mathcal{L}_{c e}(\boldsymbol{\theta})\right\|_2}) \approx \nabla_{ \boldsymbol{\theta}} \mathcal{L}_{ce}( \boldsymbol{\theta})|_{ \boldsymbol{\theta}= \boldsymbol{\theta}+r \frac{\nabla_{ \boldsymbol{\theta}} \mathcal{L}_{ce}( \boldsymbol{\theta})}{{ \|\nabla_{ \boldsymbol{\theta}}} \mathcal{L}_{ce}( \boldsymbol{\theta}) \|_2}}.
\end{equation}


The details of the algorithm process are illustrated in Algorithm~\ref{alg:GFT}.
For each iteration (line 1$\sim$9 of the Algorithm), we first obtain a mini-batch of data $\boldsymbol{B}_c$ (line 2) and the current parameters $\boldsymbol{\theta}^t$ (line 3). Based on them, we calculate the first term in equation~\eqref{equ:appro} as $\boldsymbol{g}_1$ (line 4). Then, we temporally conduct an additional optimization step as stated in equation~\eqref{equ:loss_appro_second} to approximate the second term in equation~\eqref{equ:appro}, as $\boldsymbol{g}_2$ (line 5$\sim$6). By combining the two terms with $\alpha$, we can calculate the final gradients to permanently update $\boldsymbol{\theta}^t$ (line 7$\sim$8). After $T$ iterations, we can obtain a repaired model $F_c$, which is validated effective towards audio-backdoored model in Section~\ref{sec:experiment}. 

\begin{algorithm}[h]   
  \caption{Gradient Norm-based Fine-Tuning}  
  \label{alg:GFT}  
  \begin{algorithmic}[1]
    \Require  
        Clean dataset $\mathcal{D}_c$; backdoored model $F$; the number of iterations $T$; hyper-parameters $r$ and $\alpha$;
     \Ensure The clean model $F_c$;
     \For{$t = 1$ to $T$}
     \State Get a mini-batch $\boldsymbol{B}_c$ from $\mathcal{D}_c$;
     \State Extract the parameters $\boldsymbol{\theta}^t$ from $F$;
     \State Input $\boldsymbol{B}_c$ into $F$, calculate gradients $\boldsymbol{g}_1 \leftarrow \nabla_{\boldsymbol{\theta}^t} \mathcal{L}_{c e}(\boldsymbol{\theta}^t)$;
     \State Copy the model $F$ as $F^{\prime}$, and define its parameters as $\boldsymbol{\theta}^{\prime} \leftarrow \boldsymbol{\theta}^t+r \frac{\boldsymbol{g}_1}{\left\|\boldsymbol{g}_1\right\|_2}$;
     \State Input $\boldsymbol{B}_c$ into $F^{\prime}$, calculate gradients $\boldsymbol{g}_2 \leftarrow \nabla_{\boldsymbol{\theta}^{\prime}} \mathcal{L}_{c e}(\boldsymbol{\theta}^{\prime})$;
     \State Calculate the final gradient $\boldsymbol{g}\leftarrow(1-\alpha) \boldsymbol{g}_1+\alpha \boldsymbol{g}_2$;
     \State Update $\boldsymbol{\theta}^t$ using $\boldsymbol{g}$;   
  	\EndFor
   
        \State \textbf{Return} $F_c$ with parameters $\boldsymbol{\theta}^T$.
  \end{algorithmic} 
\end{algorithm}

\section{Experiments}
\label{sec:experiment}
\subsection{Experimental Setups }
\noindent \textbf{Datasets and Models. }We use Google's Speech Commands Dataset (SCD) \cite{warden2018speech}, a commonly used dataset for speech recognition tasks. We employ two versions: the first version contains 10 classes that were also used in \cite{koffas2022can} (SCD-10), and the second version includes the full 30 classes (SCD-30). We choose ResNet \cite{he2016deep}, LSTM \cite{hochreiter1997long}, Small CNN \cite{wu2017introduction, samizade2020adversarial}, KWT \cite{berg2021keyword} and EAT \cite{gazneli2022end}, which are commonly used as speech recognition models, as our experimental models.

\noindent \textbf{Attacker Settings. }In our experiments, the poisoning ratio for the attacks is set to 10\%, and the target label is set to ``up''. We employ seven attack methods: Audio BadNets~\cite{gu2019badnets},  Ultrasonic \cite{koffas2022can}, JingleBack \cite{koffas2023going}, DABA \cite{liu2022opportunistic}, FlowMur \cite{lan2024flowmur}, PBSM and VSVC \cite{cai2024towards}. Among these, we extend the representative attack, BadNets \cite{gu2019badnets} from the visual domain, to Audio BadNets in the audio domain, where a white block is added at a fixed position in the MFCC spectrogram \cite{gupta2013feature} of the audio signal to be poisoned. 

\noindent \textbf{Defender Settings. }Since our \nameFramework is designed as a model-level defense, we compare it with the only known model-level defense method adapted to the audio domain, FP \cite{liu2018fine}. We follow a similar setting with 5\% (known as \textit{clean data ratio}) of the clean training data for defense purposes. Since the hyperparameters $r$ and $\alpha$ are more related to the approximation ability, and well-discussed in \cite{zhao2022penalizing}, we 
follow the default setup to 0.05 and 0.7, respectively.

\noindent \textbf{Evaluation Metrics. }We use two metrics to evaluate the defense methods: Clean Accuracy (\textbf{CA}) and Attack Success Rate (\textbf{ASR}). CA represents the accuracy of the model on clean data, while ASR indicates the proportion of poisoned data that the model predicts as the target label. The \textbf{boldfaced} numbers represent the best performance among the same metric. 

\subsection{Main Results}
\noindent\textbf{Results On SCD-10. }Table \ref{table_result-SCD10} presents the  performance comparisons on SCD-10 dataset using ResNet, LSTM, Small CNN, KWT and EAT. The results demonstrate that our method, \nameFramework, exhibits significant advantages over FP, effectively reducing ASR while maintaining a high CA in most cases. For ResNet, \nameFramework significantly lowers the average ASR from 91.72\% to 9.73\%, while the average CA only drops slightly from 94.54\% to 90.40\%. 
Although FP performs better ASR in JingleBack and DABA, the sacrifices on CA are unacceptable at 33.69\% and 21.45\%, respectively. 
Similarly, for LSTM, Small CNN, KWT and EAT, \nameFramework outperforms FP for nearly all attacks. In contrast, FP fails to achieve effective defense against these attack methods. 

\noindent \textbf{Results on SCD-30. }Table \ref{table_result-SCD30} presents the defense performance on SCD-30 dataset using ResNet and LSTM. Similar to its performance on SCD-10 dataset, \nameFramework significantly outperforms FP on SCD-30 dataset as well. For ResNet, \nameFramework demonstrates effective defense against most attacks, although less effective on ASR towards the more stealthy attacks, JingleBack and FlowMur. 
Notably, in DABA and FlowMur attacks, \nameFramework can increase the model’s CA  compared to No Defense, improving it by 0.07\% and 8.68\%, respectively, and the average CA after defense also increases by 0.74\%. For LSTM, \nameFramework can successfully defend against all five attacks with ASR lower than 10\%.
Similar to the performance on SCD-10 dataset, FP suffers from high ASRs or significant reduction in CAs. 

Overall, we can see that \nameFramework is an effective defense method against all attack methods under different experimental setups.

\begin{table}[t]
\captionof{table}{Main experimental results on SCD-10 dataset (\%).}
\label{table_result-SCD10}
\vspace{+2pt}
\resizebox{\linewidth}{!}{
\begin{tabular}{c|c|cc|cc|cc}
\hline
\multirow{2}{*}{Models} & \multirow{2}{*}{Backdoor attacks} & \multicolumn{2}{c|}{No Defense} & \multicolumn{2}{c|}{FP~\cite{liu2018fine}}          & \multicolumn{2}{c}{\nameFramework(Ours)}  \\ \cline{3-8} 
                        &                                   & ASR $\downarrow$           & CA  $\uparrow$        & ASR $\downarrow$           & CA $\uparrow$           & ASR $\downarrow$         & CA $\uparrow$          \\ \hline
\multirow{6}{*}{ResNet} & Audio BadNets \cite{gu2019badnets}                    & 99.27          & 92.44         & 25.20          & 44.64          & \textbf{2.92}  & \textbf{91.06} \\
                        & Ultrasonic \cite{koffas2022can}                        & 100.00         & 93.68         & 7.81           & 28.71          & \textbf{0.21}  & \textbf{90.50} \\
                        & JingleBack \cite{koffas2023going}                       & 97.42          & 92.84         & \textbf{5.44}  & 33.69          & 15.99 & \textbf{89.98} \\
                        & DABA \cite{liu2022opportunistic}                             & 99.32          & 99.91         & \textbf{0.50}  & 21.45          & 9.43  & \textbf{89.15} \\
                        & FlowMur \cite{lan2024flowmur}                          & 62.59          & 93.85         & 86.23          & 29.63          & \textbf{20.10} & \textbf{91.30} \\ \cline{2-8}
                        & Average                           & 91.72          & 94.54         & 25.04          & 31.62 & \textbf{9.73}  & \textbf{90.40} \\ \hline
\multirow{6}{*}{LSTM}   & Audio BadNets \cite{gu2019badnets}                   & 100.00         & 94.37         & 82.06          & 81.40          & \textbf{3.96}  & \textbf{85.17} \\
                        & Ultrasonic \cite{koffas2022can}                        & 100.00         & 94.37         & 97.21          & 51.42          & \textbf{1.85}  & \textbf{89.94} \\
                        & JingleBack \cite{koffas2023going}                       & 99.40          & 93.50         & 86.07          & 75.28          & \textbf{6.25}  & \textbf{85.93} \\
                        & DABA \cite{liu2022opportunistic}                             & 99.26          & 99.27         & 98.21          & 83.97          & \textbf{9.08}  & \textbf{88.04} \\
                        & FlowMur \cite{lan2024flowmur}                          & 75.60          & 92.38         & 34.30          & 62.20          & \textbf{33.10} & \textbf{86.46} \\ \cline{2-8}
                        & Average                           & 94.85          & 94.78         & 79.57          & 70.85          & \textbf{10.85}          & \textbf{87.11}          \\ \hline
\multirow{4}{*}{\begin{tabular}[c]{@{}c@{}}Small\\ CNN\end{tabular}}    & Audio BadNets \cite{gu2019badnets}                    & 100.00         & 90.12         & 69.41          & 66.71          & \textbf{25.02} & \textbf{82.03} \\
                        & Ultrasonic \cite{koffas2022can}                       & 99.97          & 91.23         & 86.33          & 69.36          & \textbf{23.17} & \textbf{78.61} \\
                        & JingleBack \cite{koffas2023going}                       & 98.88          & 90.38         & \textbf{39.36} & 53.65          & 48.32          & \textbf{81.02} \\ \cline{2-8}
                        & Average                           & 99.62          & 90.58         & 65.03          & 63.24          & \textbf{32.17}          & \textbf{80.55}          \\ \hline
\multirow{3}{*}{KWT}   & PBSM \cite{cai2024towards}                   & 92.20         & 91.47         & \textbf{17.17}          & 71.41         & 17.30  & \textbf{84.41} \\
                        & VSVC \cite{cai2024towards}                        & 99.83         & 91.16         & 23.18          & 71.60          & \textbf{15.90}  & \textbf{84.11} \\
                        \cline{2-8}
                        & Average                           & 96.02          & 91.32         & 20.18          & 71.51          & \textbf{16.60}          & \textbf{84.26}          \\ \hline
\multirow{3}{*}{EAT}   & PBSM \cite{cai2024towards}                   & 100.00        & 95.60         & \textbf{0.00}          & 10.09         & 2.92  & \textbf{94.94} \\
                        & VSVC \cite{cai2024towards}                        & 99.08         & 95.36         & \textbf{0.00}          & 9.97          & 2.79  & \textbf{95.33} \\
                        \cline{2-8}
                        & Average                           & 99.54          & 95.48         & \textbf{0.00}          & 10.03          & 2.86          & \textbf{95.14}          \\ \hline
\end{tabular}}
\captionof{table}{Main experimental results on SCD-30 dataset (\%).}
\label{table_result-SCD30}
\vspace{+2pt}
\resizebox{\linewidth}{!}{
\begin{tabular}{c|c|cc|cc|cc}
\hline
\multirow{2}{*}{Models} & \multirow{2}{*}{Backdoor attacks} & \multicolumn{2}{c|}{No Defense} & \multicolumn{2}{c|}{FP~\cite{liu2018fine}}          & \multicolumn{2}{c}{\nameFramework(Ours)}  \\ \cline{3-8} 
                        &                                   & ASR $\downarrow$           & CA  $\uparrow$          & ASR $\downarrow$           & CA  $\uparrow$            & ASR $\downarrow$            & CA $\uparrow$             \\ \hline
\multirow{6}{*}{ResNet} & Audio BadNets \cite{gu2019badnets}                  & 99.96          & 92.36         & 28.34          & 55.23          & \textbf{1.73}  & \textbf{89.54} \\
                        & Ultrasonic \cite{koffas2022can}                       & 100.00         & 91.37         & 29.68          & 55.78          & \textbf{0.12}  & \textbf{90.02} \\
                        & JingleBack \cite{koffas2023going}                       & 99.21          & 90.67         & 36.97 & 47.70          & \textbf{28.18} & \textbf{89.76} \\
                        & DABA \cite{liu2022opportunistic}                             & 99.63          & 80.09         & 71.63 & 48.97          & \textbf{4.51}  & \textbf{88.77} \\
                        & FlowMur \cite{lan2024flowmur}                          & 41.31          & 87.87         & 51.40          & 30.30          & \textbf{42.40} & \textbf{87.94} \\ \cline{2-8}
                        & Average                           & 88.02          & 88.47         & 43.60          & 47.60 & \textbf{15.39} & \textbf{89.21} \\ \hline
\multirow{6}{*}{LSTM}   & Audio BadNets \cite{gu2019badnets}                   & 100.00         & 94.04         & 83.23          & \textbf{82.59} & \textbf{1.19}  & 82.47 \\
                        & Ultrasonic \cite{koffas2022can}                       & 100.00         & 93.81         & 99.60          & 83.34          & \textbf{0.12}  & \textbf{90.02} \\
                        & JingleBack  \cite{koffas2023going}                      & 99.72          & 93.96         & 84.10          & 79.95          & \textbf{7.97}  & \textbf{84.75} \\
                        & DABA \cite{liu2022opportunistic}                             & 99.90          & 78.90         & 98.85          & 53.13          & \textbf{4.02}  & \textbf{82.74} \\
                        & FlowMur \cite{lan2024flowmur}                          & 46.93          & 91.05         & 4.18           & 66.53          & \textbf{2.72}  & \textbf{82.16} \\ \cline{2-8}
                        & Average                           & 89.31          & 90.35         & 73.99          & 73.11          & \textbf{3.20}           & \textbf{84.43}          \\ \hline
\end{tabular}}
\vspace{-1.5em}
\end{table}

\subsection{Ablation Studies}
To verify the effectiveness of gradient regularization, we compare it with the \textit{vanilla Fine-Tuning} (FT for short). As shown in Table \ref{table_ablation}, the defense performances of FT on the audio-backdoored models are limited: despite the high CAs, it fails to reduce ASRs. On the contrary, \nameFramework significantly reduces ASRs with a similar sacrifice on CAs. This further underscores the importance of gradient regularization. 

\vspace{-1em}
\begin{table}[htb]
\centering
\captionof{table}{Comparison with FT on SCD-10 using ResNet (\%). }
\label{table_ablation}
\vspace{+2pt}
\begin{tabular}{c|cc|cc}
\hline
\multicolumn{1}{c|}{\multirow{2}{*}{Backdoor attacks}} & \multicolumn{2}{c|}{FT} & \multicolumn{2}{c}{\nameFramework(Ours)} \\ \cline{2-5} 
\multicolumn{1}{c|}{}                                  & ASR$\downarrow$        & CA$\uparrow$       & ASR$\downarrow$            & CA$\uparrow$           \\ \hline
Audio BadNets \cite{gu2019badnets}                                       & 99.38      & 91.32     & 2.92           & 91.06         \\
Ultrasonic \cite{koffas2022can}                                            & 100.00        & 92.68     & 0.20            & 90.50          \\
JingleBack \cite{koffas2023going}                                            & 72.19      & 90.71     & 15.99          & 89.98         \\ \hline
\end{tabular}
\end{table}
\vspace{-1em}

\subsection{Further Analysis }

\noindent\textbf{Impact of Clean Data Ratio. }We analyze the impact of different clean data ratios, \ie, the proportion of clean data used for defense, on the defense performance. As shown in Table \ref{table_size_impact}, a smaller clean data ratio is prone to bring a worse defense performance, especially for the sacrifice of CAs. 
Once the proportion of clean data exceeds 10\%, CAs can maintain stably high at above 90\% for different attacks, and ASRs decrease to around 5\% or even lower. This indicates that the performance of \nameFramework can be further improved and stabilized if there were more clean data available for defense.

\begin{table}[t]
\captionof{table}{The impact of the clean data ratio on the defense performance using ResNet (\%).}
\label{table_size_impact}
\vspace{+2pt}
\resizebox{\linewidth}{!}{
\begin{tabular}{c|cc|cc|cc}
\hline
\multirow{2}{*}{Clean Data Ratio} & \multicolumn{2}{c|}{Audio BadNets \cite{gu2019badnets}} & \multicolumn{2}{c|}{Ultrasonic \cite{koffas2022can}} & \multicolumn{2}{c}{JingleBack \cite{koffas2023going}} \\ \cline{2-7} 
                              & ASR$\downarrow$             & CA$\uparrow$               & ASR$\downarrow$           & CA$\uparrow$             & ASR$\downarrow$            & CA$\uparrow$            \\ \hline
2\%                          & 10.00              & 85.20             & 5.86          & 85.15          & 11.66          & 85.51         \\
5\%                          & 2.92            & 91.06            & 0.21          & 90.50           & 15.99          & 89.98         \\
10\%                           & 3.54            & 93.46            & 1.04          & 92.87          & 5.26           & 92.87         \\
20\%                           & 3.31            & 95.17            & 2.47          & 94.70           & 3.64           & 94.75         \\
40\%                           & 2.88            & 95.89            & 0.76          & 95.64          & 2.66           & 95.64         \\ \hline
\end{tabular}}
\vspace{-1.5em}
\end{table}

\begin{figure}[h]
\setlength{\abovecaptionskip}{3pt}
\begin{minipage}[b]{1.0\linewidth}
 \centering
 \centerline{\includegraphics[width=5cm]{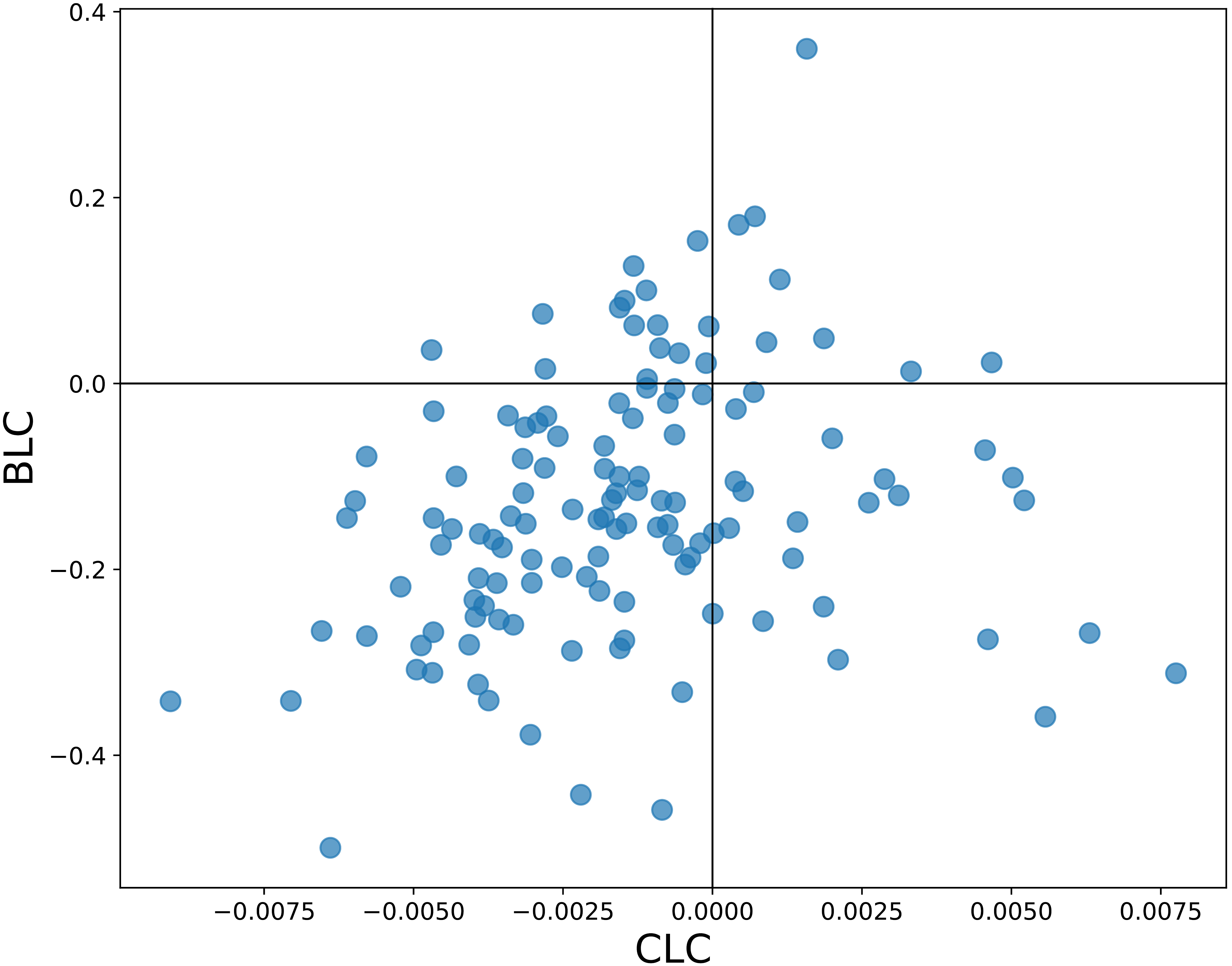}}
\end{minipage}
\caption{The BLC-CLC distribution of neurons after \nameFramework against Audio BadNets.}
\label{fig_ft}
\end{figure}

\noindent \textbf{Influences on Backdoored and Hybrid Neurons. }To analyze the influences of \nameFramework on the backdoor and hybrid neurons, we observe the BLC-CLC distribution of neurons in the last two convolutional layers after defense. As shown in Fig. \ref{fig_ft}, \nameFramework can effectively reduce the number of neurons in the H-zone and B-zone compared to Fig.~\ref{fig_example}, while the number of neurons in the R-zone increases, indicating that some hybrid neurons and backdoored neurons have been moved to the R-zone after defense, which is considered redundant. The result proves that our method can mitigate the backdoor by reducing the number of backdoored and hybrid neurons.

\vspace{-0.5em}

\begin{figure}[h]
\setlength{\abovecaptionskip}{3pt}
\begin{minipage}[b]{1.0\linewidth}
 \centering
 \centerline{\includegraphics[width=8.8cm]{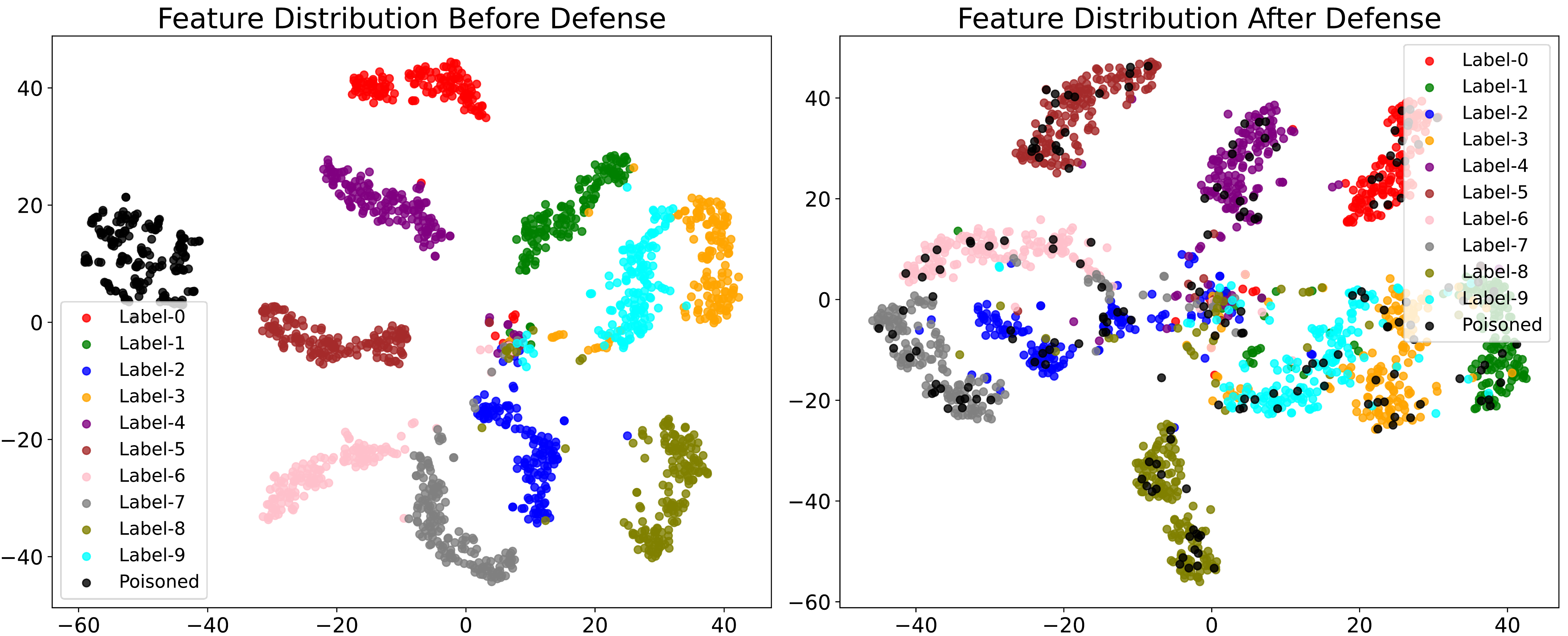}}
\end{minipage}
\caption{The t-SNE plots before and after \nameFramework against BadNets using SCD-10. \textbf{Black} points indicate the poisoned features. }
\label{fig_tsne}
\end{figure}

\noindent \textbf{T-SNE Visualization. }
Fig.\ref{fig_tsne} shows t-SNE \cite{van2008visualizing} plots before and after \nameFramework defense. Before defense, the poisoned features form a clear cluster (left of Fig.\ref{fig_tsne}), indicating the trigger features are well-learned. After defense, the poisoned features become dispersed and distributed among other classes (right of Fig.~\ref{fig_tsne}), while clean data features remain clustered. This suggests the model "forgets" the trigger features but retains performance on clean data.

\vspace{-0.5em}
\vspace{-0.5em}

\section{Conclusion}

We propose Gradient Norm-based Fine-Tuning to mitigate the backdoored effects in audio-backdoored models. Our study reveals that backdoored and hybrid neurons in these models show larger gradients for clean inputs. Our work highlights the need for specific backdoor defenses for audio-backdoored models, which differ from visual models but were previously overlooked. Our future work will focus on exploring the differences between audio and visual domains to enhance the defense performance against stealthy attacks like DABA and FlowMur.

\section{Acknowledgment}
This work was supported by the National Natural Science Foundation of China (No. 62471420 and 62101351), and Guangzhou-HKUST(GZ) Joint Funding Program (Grant No.2023A03J0008), Education Bureau of Guangzhou Municipality.

\clearpage
\bibliographystyle{IEEEbib}

\end{document}